\documentclass{iau}
\usepackage{graphicx}
\usepackage{natbib}
\usepackage{epstopdf}

\def\apjl{{\em ApJL}}
\def\aj{{\em AJ}}
\def\apj{{\em ApJ}}
\def\pasp{{\em PASP}}

\def\apjs{{\em ApJS}}

\def\aap{{\em A\&A}}

\def\mnras{{\em MNRAS}}
\def\pasj{{\em PASJ}}
\title[From Gas to Stars] 
{From Gas to Stars over Cosmic Time}

\author[Mordecai-Mark Mac Low]   
{Mordecai-Mark Mac Low$^{1,2}$
}

\affiliation{$^1$Department of Astrophysics, American Museum of
  Natural History \\ 79th Street at Central Park West, New York, NY,
  10024-5192, USA  \\ email: {\tt mordecai@amnh.org} \\[\affilskip]
$^2$Institut f\"ur Theoretische Astrophysik, Zentrum
  f\"ur Astronomie der Universit\"at Heidelberg}
\pubyear{2013}
\volume{292}  
\pagerange{}
\jname{Molecular Gas, Dust, and Star Formation in Galaxies}
\editors{T. Wong, J. Ott, eds.}
\begin{document}

\maketitle
\begin{abstract}
  The formation of stars from gas drives the evolution of
  galaxies. Yet, it remains one of the hardest processes to understand
  when trying to connect observations of modern and high-redshift
  stellar and galaxy populations to models of large scale structure
  formation.  It has become clear that the star formation rate at
  redshifts $z>2$ drops off rather more quickly than was thought even
  five years ago.  Theoretical models have tended to overpredict the
  star formation rate at these high redshifts substantially, primarily
  due to overcooling.  Overcooling in galaxies typically occurs
  because of unphysical radiative cooling.  As a result, insufficient
  turbulence is driven by stellar feedback in galaxies.  I show that
  such turbulence has the net effect of strongly inhibiting star
  formation, despite its ability to locally promote star formation by
  compression. Radiation pressure appears less likely to be a dominant
  driver of the turbulence than has been argued, but supernova and
  magnetorotational instabilities remain viable mechanisms.  Gravity
  alone cannot be the main driver, as otherwise well-resolved models
  without feedback would accurately predict star formation rates.
  Star formation rate surface density correlates well with observed
  molecular gas surface density, as well as with other tracers of high
  density material.  Correlation does not, however, necessarily imply
  causation.  In this case, it appears that both molecule formation
  and star formation occur as a consequence of gravitational collapse,
  with molecules typically playing an important but not an essential
  role in cooling.  The basic concept that gravitational instability
  drives star formation remains a true guide through the thickets of
  complexity surrounding this topic. I finally briefly note that
  understanding ionization heating and radiation pressure from the
  most massive stars will likely require much higher resolution models
  (sub-parsec scale) than resolving supernova feedback.
 
\end{abstract}

\section{Star Formation History of the Universe}
\label{SFhistory}
\begin{figure}
\begin{center}
 \includegraphics[width=0.7\textwidth]{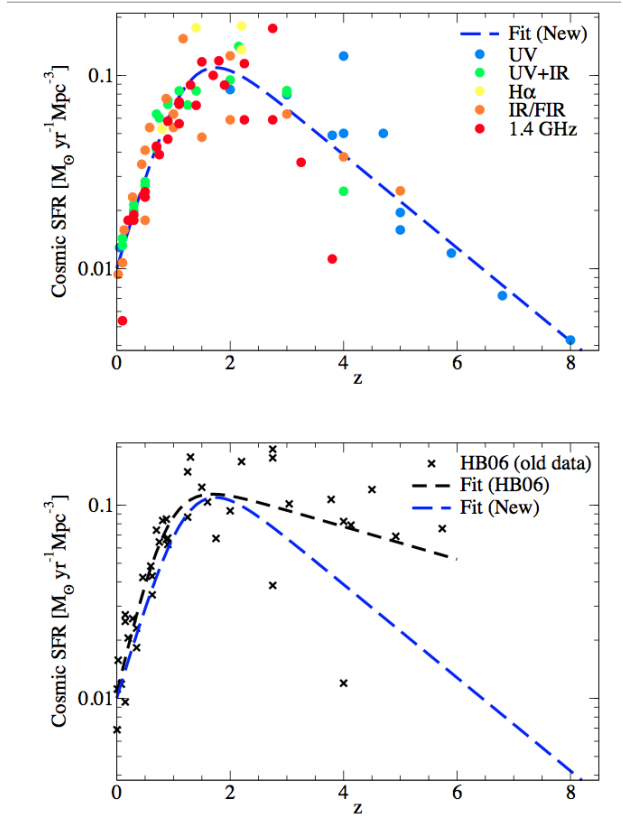}\end{center}
\caption{\label{SFR} (a) Star formation rate density as a function of
  cosmic time, showing a fit to current data coded by observation
  wavelength range. (b) A comparison to the shallower drop-off at high
  redshift found by fitting to data reported prior to 2006 shown as
  crosses \citep{hopkins2006} Figure from \citep{behroozi2012} }
\end{figure}

Five years ago, the star formation rate in the Universe was
thought to peak at redshifts $z \sim 2$--3, with a
rather shallow drop off beyond that era \citep[e.g.][]{hopkins2006}.
Recent observations from radio to (rest-frame) UV wavelengths have
reached consensus on the star formation 
history of the Universe dropping rather faster than previously thought
\citep[e.g.][]{behroozi2012,moster2012}, as shown in
Figure~\ref{SFR}. 

The discrepancies between the older and more recent measurements
appear to be dominated by two effects \citep{reddy2009}.  First, dust
corrections to star formation rates derived from rest-frame UV
emission probably do not remain constant beyond redshift $z=2$, but
rather drop at higher redshift and for lower luminosity galaxies,
which likely have substantially lower metallicities and thus dust
abundances.  Second, the faint end of the UV luminosity function may
be rather steeper than previously thought.

The star formation rate density is dominated at all
redshifts by galaxies with stellar masses of a few $\times
10^{10}$~M$_{\odot}$ \citep{karim2011}, comparable to modern irregular
galaxies such as the Large Magellanic Cloud.  However, high redshift
galaxies in this mass range had far higher accretion and star
formation rates than modern galaxies of similar masses, so analogies
between them cannot always be drawn.  Such high-redshift galaxies have
by $z=0$ typically evolved into much more massive galaxies.

\section{Overcooling}
\label{overcooling}
Theoretical models a decade ago had predicted a rather earlier peak in
star formation, at $z \sim 5$--6 \citep[e.g.][]{springel2003}.  This
contradiction to the observations reflects a fundamental issue in
cosmological models of star formation, that simulations either
overproduce stars at early times \citep{white1991}, or rely on ad hoc
models of strong feedback to suppress that early star formation, in
order to agree with the observations.  There are two reasons for this
requirement.  First, accretion onto massive elliptical galaxies is prevented,
perhaps by AGN feedback.  It is much easier to prevent accretion of
diffuse gas that can not cool easily, than it is to reheat and expel
already accreted gas.  I will not further discuss this aspect of the
problem in this contribution, though.  

Second, simulations capturing cosmological scales have been unable to
model the interstellar medium with sufficient resolution to follow the
energetics of stellar feedback successfully, leading to the classical
overcooling problem.  Without sufficiently energetic local feedback,
the star formation rate can be an order of magnitude higher than
observed, even in models of modern galaxies \citep{tasker2011}, a
conclusion also reached by many previous and current workers,
including \citet{katz1996,somerville1999,cole2000,
  springel2003,keres2009,bournaud2010, dobbs2011}, and
\citet{hummels2012}.

Feedback models typically fail because of unphysical cooling.  The
fundamental problems are that the radiative cooling rate of gas in
ionization equilibrium $\dot{E} = -n^2 \Lambda(T)$ depends nonlinearly
on density, and $\Lambda(T)$ is more than an order of magnitude higher
for $T = 10^5$~K gas than for hot $10^6$~K or cool $10^4$~K gas
\citep{sutherland1993}.  The elevated cooling around $10^5$~K occurs
because the strong resonance lines of lithium-like ions of the most
common metals carbon, oxygen (and nitrogen) can be excited.

These properties of the cooling function lead to two problems for
numerical models.  First, if feedback energy is fed into the gas too
slowly or over too large a volume, it will only raise the temperature
into the $10^5$~K range, so that the energy will be promptly be lost
to radiation without exerting dynamical effects.  Real supernova
remnants, on the other hand, produce gas hotter than $10^6$~K that
only cools with difficulty.  Second, in poorly resolved models cool
dense gas can numerically diffuse across interfaces with hot, rarefied
gas.  This can produce large volumes of gas subject to unphysically
strong radiative cooling, because they are elevated in density and
reduced in temperature compared to the physical solution.  Already
over 25 years ago, \citet{tomisaka1986} demonstrated that the evolution
of superbubbles formed by multiple supernova explosions could not be
adequately followed with 5~pc resolution because of such strong numerical
overcooling.

For models of the diffuse interstellar medium (ISM) in the Milky Way
($0.1 < n < 100$~cm$^{-3}$), models by \citet{avillez2000, joung2006,
  hill2012} and others have demonstrated that 2~pc resolution is
generally sufficient to resolve interfaces sufficiently to avoid
dynamically important loss of energy from hot gas.  Models of modern
dwarf galaxies at lower average densities can tolerate reduced
resolution, as even the dense, swept up supershells have lower
densities, and thus induce less cooling in the hot gas
\citep{fragile2004}.

\section{Turbulent Inhibition of Star Formation}
\label{turbSF}
Highly compressible turbulence driven by supernovae and other feedback
mechanisms both promotes and prevents
gravitational collapse.  We can estimate which effect is more
important by examining the dependence of
the Jeans mass  
\begin{equation} M_{\rm J} \propto \rho^{-1/2} c_{\rm s}^3
\end{equation}
on the rms turbulent velocity $v_{\rm rms}$ \citep{mac-low2004}.  If
we follow the classical picture that treats turbulence as an
additional pressure \citep{chandrasekhar1951,von-weizsaecker1951},
then we can define an effective sound speed $c_{{\rm s,eff}}^2 =
c_{\rm s}^2 + v_{\rm rms}^2/3$. This increases the Jeans mass by
$M_{\rm J} \propto v_{\rm rms}^3$, inhibiting collapse.  On the other
hand, shock waves with Mach number ${\cal M} = v_s / c_s$ in an
isothermal medium cause density enhancements $\rho_s / \rho_0 = {\cal
  M}^2$. Thus supersonic turbulent compression decreases the Jeans
mass by $M_J \propto \rho_s^{-1/2}$, if we assume that the shocks
typically have $v_s \simeq v_{\rm rms}$.

When we combine these two effects, we find that
\begin{equation}
M_{\rm J} \propto \left(\frac{c_s}{v_{\rm rms}}\right) \left(c_s^2 +
  \frac{v_{\rm rms}^2}{3}\right)^{3/2} \propto v_{\rm rms}^2
\end{equation}
for $v_{\rm rms} \gg c_{\rm s}$. Thus, turbulence strongly inhibits
collapse. Because it is intermittent however, even though Its net
effect is to inhibit collapse globally, it can still promote it
locally, in shock compressed regions.  A region that does not
exceed the turbulent Jeans mass globally can therefore still display some
gravitational collapse, but at low efficiency \citep{klessen2000}.

This effect can also be demonstrated in the diffuse, stratified
interstellar medium.  \citet{joung2006} used the Flash adaptive mesh
refinement code \citep{fryxell2000} to run well-resolved models of
supernova driving of turbulence in the ISM, including heating and
cooling, but not self-gravity.  They indeed found Jeans-unstable
regions of cold, dense gas with sizes comparable to observed molecular
clouds. However, if the star formation rate expected for those regions
is computed, it is an order of magnitude below the rate required to
produce the assumed supernova driving.  Triggering of star formation
by turbulence only occurs at low efficiency, and cannot lead to
stochastic propagation waves as once envisioned by
\citet{elmegreen1977}.

At smaller scales, \citet{dale2007} modeled the effect of ionizing
radiation on a turbulent molecular cloud.  The morphology of the cloud
was drastically modified, as the radiation ionized and heated
low-density gas that expanded outwards, driving compressive shock
waves into the surrounding cloud.  However, the actual difference in
the star formation rate was small, with the net effect being to
accelerate star formation by perhaps $0.2 t_{\rm ff}$, where the
free-fall time $t_{\rm ff} = (3\pi/32 G \rho)^{1/2}$.

Quantitative observational studies reveal results consistent with this
description. Although triggered star formation clearly occurs, it is a
relatively small effect that does not explain most star formation. For
example, \citet{getman2012} show that, even under favorable
circumstances, less than a quarter of star formation in the Elephant
Trunk Nebula is due to triggered star formation. At the galactic
scale, in the Large Magellanic Cloud, supergiant shells represent the
largest-scale compressive structures in that galaxy. However, they
contain only about 10\% of young clusters \citep{yamaguchi2001} and
about 5\% of the molecular gas (Dawson et al., this volume).

\section{Sources of Turbulence}
\label{sources}
If turbulence controls star formation, then understanding the sources
of turbulence, both in the diffuse ISM, and in molecular clouds, will
help us to understand star formation. 

Recently, radiation pressure from the most massive star clusters has
been argued to play a dominant role in limiting star formation by a
number of groups including
\citet{thompson2005,murray2010,andrews2011}, and \citet{hopkins2011}.
However, this conclusion depends on how well radiation pressure can
couple to gas motions. If each photon only scatters once off of a gas
particle, then the strength of the radiative driving from a cluster
with luminosity $L$ is proportional to $L/c$, which is sometimes
called the momentum-driven limit.  If on the other hand the gas is
extremely optically thick, so that photons continue scattering until
they lose almost all their energy, then the driving is far higher,
proportional to $L/v_{\rm rms}$, sometimes called the energy-driven
limit.  Although it is unlikely that this limit is ever reached in
star-forming galaxies, the groups mentioned above have argued that it is
realistic to expect the number of times photons scatter to be
comparable to the infrared optical depth $\tau_{\rm IR}$, which can be
substantial.  This leads to a strength proportional to $\tau_{\rm IR}
L / c$.  On the other hand, \citet{krumholz2009} and \citet{fall2010}
argued that the momentum-driven limit was more appropriate, leading to
radiation pressure being far less important in galactic evolution.

\citet{krumholz2012a} performed multi-dimensional simulations of
radiation pressure acting on an optically thick layer of gas with
optical depth $\tau_{\rm IR} \gg 1$ to resolve this question.  As had
already been noted in models of individual massive stars
\citep{krumholz2009a}, the radiation acts as a light fluid
accelerating a heavy fluid, and thus a radiatively driven flow is
subject to Rayleigh-Taylor instability.  This overturns and fragments
the gas, stirring it, but allowing the radiation to escape far more
quickly than would be expected from its initial optical depth.  They find
that, although radiation pressure driving is indeed somewhat more
efficient than in the momentum-driven limit, it is typically at least
an order of magnitude less efficient than $\tau_{\rm IR} L/c$, calling
into serious question results based on that assumption.

The effect of supernova feedback on the diffuse ISM as the supernova
rate varies from the Milky Way value to starburst levels of as much as
512 times higher was studied by \citet{joung2009}. They varied the
midplane gas surface density with the supernova rate following the
\citet{kennicutt1998} relation between surface density and star
formation rate.  They found that regardless of surface density, the
supernovae drove a rather uniform velocity dispersion $v_{\rm rms} =
5$--10~km~s$^{-1}$, with associated H~{\sc i} linewidths of
10--20~km~s$^{-1}$ if single Gaussian components are fit to gas in the
atomic temperature range. This agrees with the vast majority of
observations of galaxies \citep[e.g.][]{petric2007,tamburro2009},
aside from extreme starbursts where elevated H~{\sc i} linewidths are
observed, possibly from radiation pressure driving \citep[][but see
comments above]{murray2010}.

The driving of turbulence in the ISM of a sample of nearby galaxies
observed by the THINGS \citep{THINGS} and SINGS \citep{SINGS} surveys
was studied by \citet{tamburro2009}. They found that the energy input from
supernova driving was sufficient to explain the observed kinetic
energy density of the ISM within the star-forming region of disks.
However, in outer disks, where star formation drops off strongly, they
found that some other mechanism was required.  Magnetorotational
instability (MRI) was shown by \citet{sellwood1999} to be able to
drive substantial turbulence in galactic disks.
\citet{piontek2004,piontek2005,piontek2007} used simulations to
demonstrate that velocity dispersions of the observed magnitudes could
be reached if thermal instability allowed a two-phase medium to form.
\citet{tamburro2009} in turn showed that the energy input expected
from MRI was sufficient to explain the kinetic energy seen in outer
disks of galaxies.  On the other hand, \citet{elmegreen1994} and
\citet{schaye2004} argue that the transition from a single-phase to a
two-phase medium marks the point at which ultraviolet heating can no
longer maintain the observed velocity dispersion in outer disks.
These two models can be observationally distinguished by the presence
or absence of low temperature gas in outer disks.  The discovery of
finite rates of star formation in these regions by GALEX
\citep{boissier2007}, however, seems to lean toward the presence of
a two-phase medium, supporting the MRI model.

Gravity itself can drive turbulence even in the absence of other
energy inputs.  \citet{bournaud2010} demonstrated that gravitationally
driven turbulence in unstable disks produces a column density
fluctuation power spectrum consistent with observations of neutral gas
in the Magellanic clouds.  However, because turbulence decays in a
free-fall time \citep{stone1998,mac-low1998,mac-low1999a}, such
internal gravitational turbulence cannot effectively delay star
formation on its own. 

On the other hand, accretion from the
intergalactic medium onto
galaxies brings substantial energy with it.  \citet{klessen2010}
demonstrated that if that energy couples to the ISM with only 10\%
efficiency, the velocity dispersion of spiral galaxies could be
supported if they accrete gas at the same rate that they form stars.  However,
dwarf galaxies, which sit in shallower potential wells, and thus have
lower energy accretion flows, cannot be explained by this mechanism.
As such dwarf galaxies represent the dominant location for star
formation \citep{karim2011}, other mechanisms must play an important
role in its regulation.

At the molecular cloud scale, accretion from the surrounding
interstellar medium can also play an important
role in driving observed turbulent motions
\citep{klessen2010,vazquez-semadeni2010,goldbaum2011}.  This results in
extended lifetimes compared to isolated clouds.  However, other forms
of feedback appear necessary to explain the termination of star
formation and the disruption of the dense clouds.

Models of galaxy formation do lead us to one simple conclusion:
driving of turbulence by either gravitational or accretional sources
must be supplemented by other energy sources.  Otherwise, simulations
without stellar feedback or other energy sources beyond gravity would
be sufficient to reproduce observed galaxies.  This does not appear to
be an effect of insufficient numerical resolution, as workers such as
\citet{bournaud2010} have used adaptive mesh techniques to model large
ranges of spatial scale without changing this fundamental result.
Ultimately, gravity must compete with feedback to determine collapse
and star formation.

\section{Star Formation Laws}
\label{SFlaws}
\begin{figure}
  \begin{center}
   \includegraphics[width=3in]{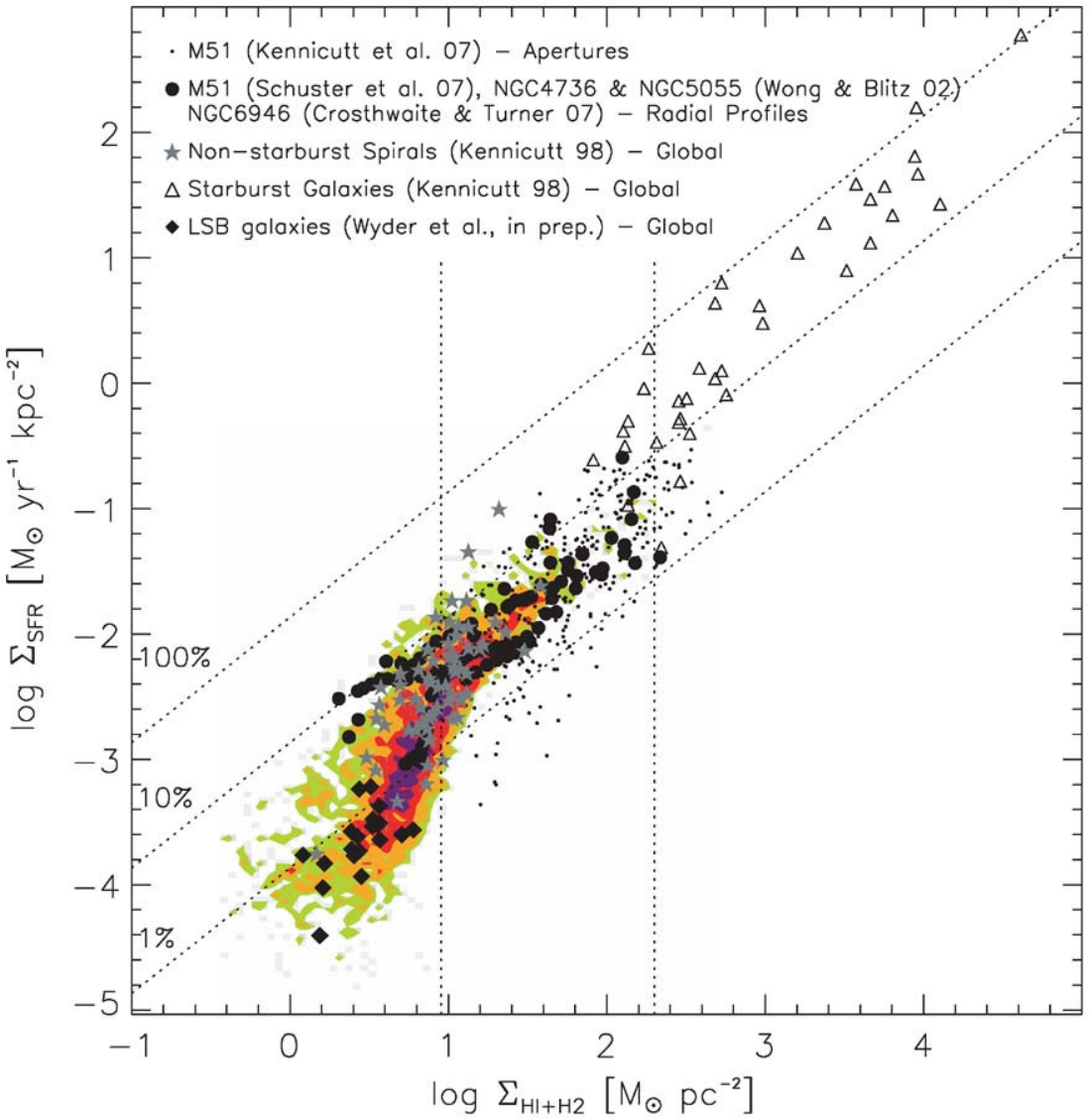}
    \includegraphics[width=3in]{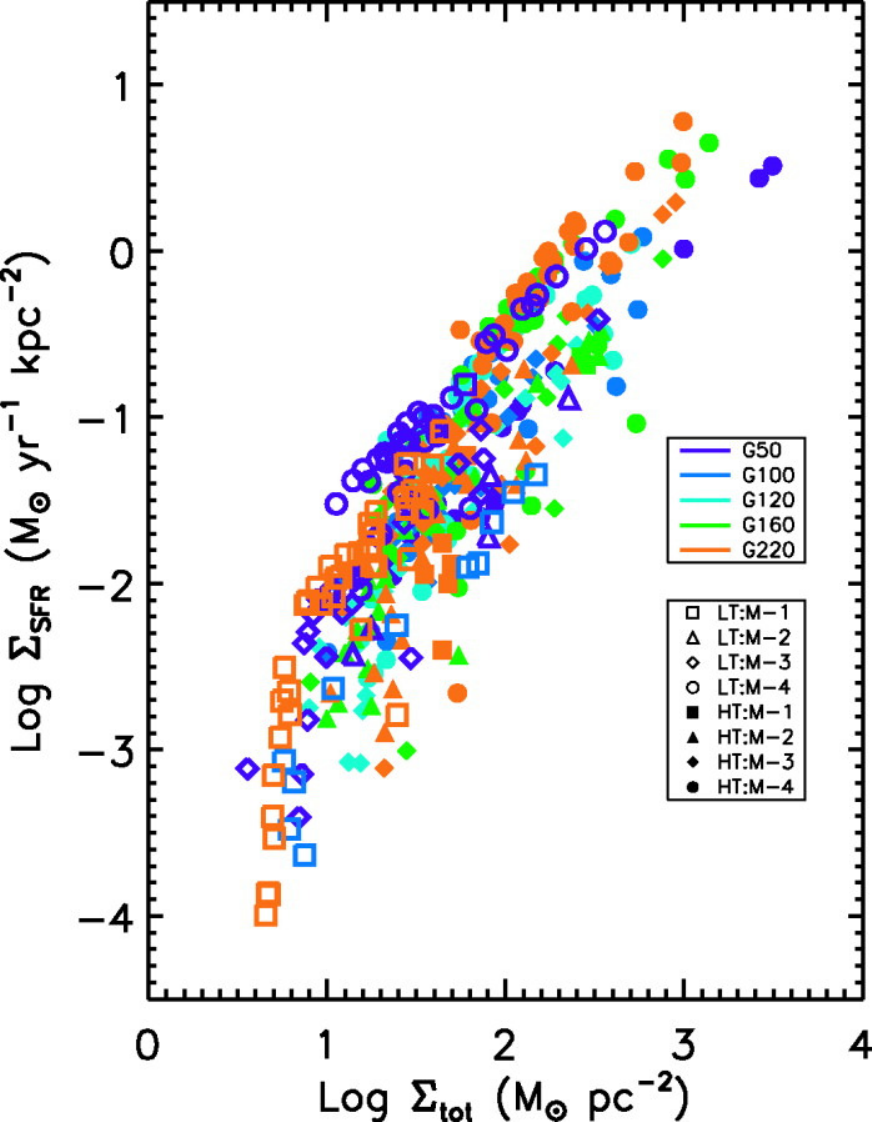}
   \end{center}
\caption{\label{pixelks} {\em (a)} A comparison of $\Sigma_{\rm SFR}$ to
  $\Sigma_{\rm gas}$ from Figure~15 of \citet{bigiel2008} showing
  combined data from that paper in colored contours, along with points
  from the observations described in the legend on the figure.  The
  dashed lines show what percentage of the gas would be consumed at that
  star formation rate over a period of $10^8$~yr. {\em (b)} Radial
  profiles across model disks simulated with isothermal gas and live
  stellar disks and dark matter halos \citep{li2005a}, showing the
  same drop in star formation efficiency at low gas surface density.}
\end{figure}
 Star formation correlates extraordinarily well with certain properties
of galaxies.  Perhaps the most well-known of these correlations is the
Kennicutt-Schmidt law relating the gas surface density $\Sigma_{\rm
  gas}$ averaged over whole disks of normal or starburst galaxies, or
entire galactic centers, to the star formation rate surface density
$\Sigma_{\rm SFR} \propto \Sigma_{\rm gas}^{1.4}$ \citep{kennicutt1998}.  As
observational resolution has improved a similar correlation has now
been found for regions of size around a square kiloparsec within
galaxies \citep{bigiel2008}. However, as shown in
Figure~\ref{pixelks}{\em (a)}, very low surface density regions have
lower than expected star formation rates, while very high surface
density regions have higher than expected rates in comparison to
normal galactic star forming regions.  It is further worth noting that these
relations break down below the kiloparsec scale, with local Galactic
star-forming regions showing much more efficient star formation
\citep{heiderman2010}. 

The star formation rates in these analyses are determined using many
different indicators. The most important of these include far infrared
emission tracing deeply embedded star formation; H$\alpha$ emission
tracing emerging H~{\sc ii} regions; and far ultraviolet emission
tracing young, massive stars that have dispersed their natal gas and
dust. Different regions emit strongly in different tracers, depending on
their stage of development.  

The determination of the gas surface density depends on observation of
both atomic and molecular hydrogen surface densities.  The latter is
usually measured by observation of CO emission, followed by the use of
a calibrated conversion factor between CO luminosity and H$_2$ column
density, usually denoted $X_{\rm CO}$.  This conversion factor rises
sharply in regions with low extinction due to either low metallicity
or low column density
\citep{glover2011,narayanan2011,schruba2011,shetty2011a}.  Detailed
analysis shows that it also may drop in high column-density regions
dominated by molecular gas \citep{narayanan2012} or vary strongly when
linewidths increase \citep{shetty2011}.

Molecular hydrogen surface density correlates linearly with star
formation rate over the entire range of observed surface densities,
albeit with more than an order of magnitude scatter among individual
regions \citep{rownd1999,wong2002,bigiel2008,leroy2008,bigiel2011}.
This has been interpreted to mean that molecular hydrogen formation
controls star formation
\citep[e.g.][]{robertson2008,krumholz2009,gnedin2010,krumholz2012c,
  christensen2012}.  However, one must ask whether
correlation implies causation.

Indeed, other measurements of high density gas ($n>10^4$~cm$^{-3}$)
also yield linear correlations with star formation.  \citet{gao2004}
demonstrated that observations of HCN emission linearly correlate with
$\Sigma_{\rm SFR}$, while \citet{lada2010} showed a direct correlation
between the number of young stellar objects in a region and the mass
of material with K-band extinction $A_K > 0.8$.

The reason for this is that molecules do not appear to control star
formation in the presence of metals. Although molecular gas in fact
dominates cooling of high density gas, this is coincidental: pure
atomic gas at the same densities can cool virtually as effectively so
long as it contains even small amounts of metals.  Analytic models
\citep{krumholz2011} and numerical experiments \citep{glover2012}
demonstrate that the key to effective cooling is not molecule
formation, but rather dust shielding from photoelectric heating.
Removing molecular cooling from the models changes the minimum
temperature from 5~K to 7~K, while removing shielding increases the
minimum temperature by an order of magnitude or more
\citep{glover2012}. Indeed, \citet{hopkins2011} demonstrates that
well-resolved (parsec kernel size) galaxy formation models produce the
same result whether star formation is limited to only occur in
molecular gas, or is allowed to occur in all dense gas.

Molecular hydrogen formation occurs quickly at high density
\citep{glover2007}, so as stars form through gravitational collapse,
molecules inevitably form. \citet{glover2012a} demonstrated that this
happens almost independent of metallicity, even though molecule
formation depends on metallicity.  At solar metallicity, collapse
occurs within a free-fall time \citep{krumholz2012}. However, cooling
occurs even more quickly, within a free-fall time even for gas at
metallicity $Z > 10^{-4}$~Z$_{\odot}$.  \citet{krumholz2012}
demonstrates that in such low-metallicity gas, cooling occurs within a
free-fall time, but molecular hydrogen formation is delayed so
severely that star-formation proceeds with molecule formation only
occurring in the very densest core of the collapsing region, leading
to low integrated molecular fractions despite ongoing star
formation.

\section{Gravitational Instability}
\label{GI}
I hypothesize that gravitational instability controls the rate of star
formation in galaxies.
We can heuristically derive the \citet{toomre1964} criterion for stability
of a rotating, thin disk with uniform velocity dispersion $\sigma$ and
surface density $\Sigma$ using time scale arguments
\citep{schaye2004}, as described in \citet{mac-low2004}.
The Jeans criterion for instability in a thin disk,
requires that the time scale for collapse of a perturbation of
size $\lambda$
\begin{equation}
t_{\rm coll} = \sqrt{\lambda / G \Sigma}
\end{equation}
be shorter than the time required for the gas to respond to the
collapse, the sound crossing time
\begin{equation}
t_{\rm sc} = \lambda / c_{\rm s}.
\end{equation}
This implies that gravitational stability requires perturbations
with size
\begin{equation} \label{eqn:press}
\lambda < c_{\rm s}^2 / G \Sigma.
\end{equation}
Similarly, in a disk rotating differentially, a perturbation will
rotate around itself, generating centrifugal motions that can also
support against gravitational collapse.  This will be effective if the
collapse time scale $t_{\rm coll}$ exceeds the rotational period
$t_{\rm rot} = 2\pi/\kappa$, where $\kappa$ is the epicyclic
frequency, so that stable perturbations have
\begin{equation} \label{eqn:rot}
\lambda > 4 \pi^2 G \Sigma / \kappa^2.
\end{equation}
Gravitational instability occurs if there are
wavelengths that lie between the regimes of pressure and rotational
support, with 
\begin{equation}
\frac{c_{\rm s}^2}{G\Sigma} < \lambda < \frac{4 \pi^2 G\Sigma}{\kappa^2}.
\end{equation}
This will occur if 
\begin{equation} \label{eqn:Toomre}
Q = c_{\rm s}\kappa / (2\pi G \Sigma) < 1,
\end{equation}
which is the Toomre criterion for gravitational instability to within
a factor of two.  The full criterion from a linear analysis of the
equations of motion of gas in a shearing disk gives a factor of $\pi$
rather than $2\pi$ in the denominator \citep{safronov1960,
  goldreich1965}, while a kinetic theory approach appropriate for a
collisionless stellar system gives a factor of 3.36
\citep{toomre1964}.

When collisionless stars and collisional gas both contribute to
gravitational instability, a rather more complicated formalism is
required to accurately capture their combined action
\citep{gammie1992,rafikov2001}.  This has been, it should be noted,
successfully approximated with simple algebraic combinations of the stellar and
gas Toomre parameters \citep{wang1994,romeo2011}.  In the presence of
turbulent dissipation, \citet{elmegreen2011} demonstrates that there
is no longer a formal minimum wavelength for gravitational collapse,
although finite disk thickness does act to stabilize the smallest
wavelengths against collapse.

The relationship between global gravitational instability and star
formation can be seen in numerical experiments.  For example,
\citet{li2006} used models of isothermal, exponential gas disks
embedded in live stellar disks and dark matter halos, with gas
temperatures fixed near $10^4$~K, to study gravitational collapse as
the strength of the gravitational instability varied.  They fully
resolved the Jeans length during collapse up to pressures $P/k =
10^7$~cm$^{-3}$~K$^{-1}$, and then used sink particles to measure the
amount of gas reaching these high densities.  This required kernel
sizes less than 40~pc and several million particles.  They found not
only that all their models fell cleanly on the global correlation of
\citet{kennicutt1998}, but also, as shown in Figure~\ref{pixelks},
that an analysis of azimuthal rings predicts the behavior of the local
correlation observed by \citet{bigiel2008}.

Another example of the strength of this hypothesis lies in the
understanding of the unusual morphologies of many high-redshift
galaxies.  \citet{elmegreen2009} shows that clumpy, irregular galaxies
are far more common at redshifts $z \sim 2$ than in modern times.
Because accretion rates were far higher then, galaxies tended to be
far more gas-rich than now, and as a result were more likely to be
strongly gravitationally unstable.  \citet{agertz2009} were one group
that used well-resolved adaptive mesh computations
to show that such conditions naturally lead to the formation of giant,
gravitationally bound clumps.

An extended version of the hypothesis has been put forward by
\citet{ostriker2010}, and developed in subsequent papers
\citep{ostriker2011,kim2011}. They argue that star formation is
controlled by the combination of gravitational instability and thermal
equilibrium, because feedback increases as instability gets stronger,
until it heats the gas sufficiently to reduce the amount of
gravitationally unstable gas enough to reach a steady state.

Although resolving feedback in models reaching cosmological scales
remains tremendously difficult, computers have become sufficiently
powerful and algorithms well developed enough for this to fall
within the realm of the possible.  \citet{hopkins2011} present one
early example of this ability, with SPH models having minimum kernel
resolution of only 1~pc.  This appears to be sufficient, even with the
enhanced numerical diffusivity introduced by the SPH algorithm
\citep{bauer2012}, to maintain the energy of the hot gas and allow a
dynamically realistic interstellar medium to form.  These models do
use rather stronger radiative pressure feedback than would be
recommended by \citet{krumholz2012a}, but comparison with models
completely without radiative pressure suggests that this does not
represent a significant error. A major result from these models is
that star formation rates agreeing with observed values finally seem
to be within reach.

\section{Small Scales}
\label{small}
The conclusions drawn here have mostly focused on the effects of
feedback at scales larger than a few parsecs.  This is appropriate for
supernova feedback, because most supernovae occur far in time and
space from the dense molecular gas in which they formed: because of
the steepness of the initial mass function \citep{salpeter1955}, the vast
majority of Type II supernovae have B star progenitors with lifetimes
of 10--40~Myr.  However, this is not the case for either ionization or
radiation pressure, both of which come predominantly from the very
most massive stars, with lifetimes of only a few million years: an O4
star, for example, has ten times the ionizing luminosity of an O8
star, and 500 times that of a B0 star \citep{vacca1996}.  To
understand the effects of these processes on the diffuse ISM, small
scale models that capture the interaction of radiation with molecular
gas on sub-parsec scales, such as those by \citet{peters2010}, will
need to be included either directly, or as sub-grid scale models.

\section{Summary}

In this talk I have discussed how star formation proceeds over cosmic
time.  I began with the observational evidence that has accumulated
over the last five years demonstrating that at redshifts $z > 2$, star
formation drops off far more steeply with redshift than had been
thought previously (Sect.~\ref{SFhistory}). Theoretical predictions of
substantially higher star formation rates at high $z$ seem to have
been due to overcooling in small galaxies due to poorly modeled
feedback, as well as a lack of quenching of cooling flows onto the
most massive galaxies. Focusing on the former problem, I explained why
modeling of feedback requires that high temperature ($T > 10^6$~K) gas
be resolved with sufficient numerical resolution to avoid artificial
loss of energy through unphysical radiative cooling
(Sect.~\ref{overcooling}).

The importance of feedback comes primarily because it drives
turbulence that, on average, strongly inhibits star formation
(Sect.~\ref{turbSF}). Although it can locally trigger star formation,
this is only a 10--20\% effect in the best of cases.  The turbulence
observed in both galaxies and molecular clouds has many possible
sources (Sect.~\ref{sources}). One that has received much recent
interest, radiative pressure, may be less effective than first thought
because of Rayleigh-Taylor instabilities. Although gravity clearly can
play a role in driving the turbulence, it must ultimately be
supplemented by other sources.  Otherwise, models without or with
ineffective stellar feedback would be sufficient to reproduce observed
star formation rates.

Observed correlations between gas surface densities and star formation
rate can help us to understand how star formation proceeds.  However,
we must remember that correlation does not prove causation.  In
Sect.~\ref{SFlaws} I give evidence that the strong correlation between
the surface density of H$_2$ and star formation occurs because both
trace dense gas, rather than because the formation of H$_2$ must occur
prior to star formation. The key instead appears to be that enough
dust shielding must be present to prevent photoelectric heating and
allow cooling of the gas to around 10~K.  I then argue in
Sect.~\ref{GI} that gravitational instability controls the amount of
dense gas, and thus of star formation in galaxies.

Finally, in Sect.~\ref{small} I note the complications to be found
below the parsec scale.  Although they do not affect supernova-driven
feedback strongly, they do matter for ionization and radiation
pressure, as those are dominated by the very most massive stars, which
finish their lifetimes prior to dissipation of their natal dense gas
clouds.

\acknowledgments I thank the organizers for their kind invitation to
present this talk. I have benefited over the past decade from
discussions and collaborations on these topics with Miguel de Avillez,
Bruce Elmegreen, Simon Glover, Alex Hill, Ryan Joung, Ralf Klessen,
Mark Krumholz, Yuexing Li, and Thomas Peters. I thank Eva Schinnerer
and Joanne Dawson for providing data described here in advance of
publication, and Simon Glover for a careful reading of a draft. This
work was supported by US National Science Foundation grant
AST11-09395, Deutsche Forschungsgemeinschaft Sonderforschungsbereich 
881---The Milky Way System, and a travel grant from the International
Astronomical Union.


\end{document}